\documentclass[twocolumn]{aastex631}
\usepackage{amsmath}

\shorttitle{CDM and SIDM Interpretations of JWST-ER1}

\shortauthors{Kong, Yang, \& Yu}

\begin{document}

\title{CDM and SIDM Interpretations of the Strong Gravitational Lensing Object JWST-ER1}

\author[0000-0003-1723-8691]{Demao Kong}
\affiliation{Department of Physics and Astronomy, University of California, Riverside, California 92521, USA}

\author[0000-0002-5421-3138]{Daneng Yang}
\affiliation{Department of Physics and Astronomy, University of California, Riverside, California 92521, USA}

\author[0000-0002-8421-8597]{Hai-Bo Yu}
\affiliation{Department of Physics and Astronomy, University of California, Riverside, California 92521, USA}

\email{dkong012@ucr.edu}
\email{danengy@ucr.edu}
\email{haiboyu@ucr.edu}

\begin{abstract}

\cite{2023NatAs.tmp....7V} reported the discovery of JWST-ER1, a strong lensing object at redshift $z\approx2$, using data from the James Webb Space Telescope. The lens mass within the Einstein ring is $5.9$ times higher than the expected stellar mass from a Chabrier initial mass function, indicating a high dark matter density. In this work, we show that a cold dark matter halo, influenced by gas-driven adiabatic contraction, can account for the observed lens mass. {We interpret the measurement of JWST-ER1 in the self-interacting dark matter scenario and show that the cross section per particle mass $\sigma/m\approx0.1~{\rm cm^2/g}$ is generally favored. Intriguingly, $\sigma/m\approx0.1~{\rm cm^2/g}$ can also be consistent with the strong lensing observations of early-type galaxies at redshift $z\approx0.2$, where adiabatic contraction is not observed overall.}

\end{abstract}

\keywords{\href{http://astrothesaurus.org/uat/353}{Dark matter (353)};
\href{http://astrothesaurus.org/uat/261}{Strong gravitational lensing (261)};
\href{http://astrothesaurus.org/uat/429}{Early-type galaxies (429)};
\href{http://astrothesaurus.org/uat/1880}{Galaxy dark matter halos (1880)}}

\section{Introduction}
\label{sec:intro}

Recently,~\cite{2023NatAs.tmp....7V} used data from the James Webb Space Telescope COSMOS-Web survey and found a strong lensing object JWST-ER1 at redshift $z\approx2$. The object consists of a complete Einstein ring with a radius of $6.6~{\rm kpc}$ and a compact early-type galaxy (JWST-ER1g) with an effective radius of $1.9~{\rm kpc}$. Within the ring, the total mass is $6.5\times10^{11}~M_\odot$, while the stellar mass mass is $1.1\times10^{11}~M_\odot$ assuming a Chabrier initial mass function~\citep{2003PASP..115..763C}. The mass gap is large, indicating that JWST-ER1g has a dense halo. However, for a typical Navarro-Frenk-White (NFW) halo~\citep{Navarro_1997}, the dark matter density is not high enough to fully account for the mass gap, unless the halo mass is $10^{14}~M_\odot$~\citep{2023NatAs.tmp....7V}, which would be rare at $z\sim2$, significantly higher than expected in the stellar mass-halo mass relation~\citep{2013ApJ...770...57B}.  

\cite{2023arXiv230915986M} also analyzed the JWST-ER1 system and found the total mass is $3.7\times10^{11}~M_\odot$ within the Einstein ring, a factor of $2$ smaller than that in~\cite{2023NatAs.tmp....7V}, and a $10^{13}~M_\odot$ halo is favored for JWST-ER1g. The main reason is that their redshift measurements of the background source are different: $z\approx2.98$ in~\cite{2023NatAs.tmp....7V}, whereas $z\approx5.48$ in~\cite{2023arXiv230915986M}; their other measurements are well consistent. 

Another complication is that the population of low-mass stars could be higher than expected. For example, Salpeter-like initial mass functions can give rise to a good fit~\citep{2023NatAs.tmp....7V}, though the bottom-light Chabrier form is overall favored for quiescent galaxies at $z\gtrsim2$~\citep{2021ApJ...908L..35E,Belli:2014cra}. While more work is needed to further improve our understanding of the JWST-ER1 system, we explore its rich implications for probing dark matter physics. 

In this work, we interpret the observations of JWST-ER1 in collisionless cold dark matter (CDM) and self-interacting dark matter (SIDM) scenarios. In CDM, the halo can become denser due to adiabatic contraction induced by the infall and condensation of baryons~\citep{1986ApJ...301...27B,Gnedin_2004}. We will show that the contraction effect on the halo is beyond the stellar effective radius and it is important for the interpretation of JWST-ER1. After including adiabatic contraction in modeling the halo of JWST-ER1g, the favored halo mass is reduced to $3\times10^{13}~M_\odot$, to be consistent with the measurement in~\cite{2023NatAs.tmp....7V}; or $6\times10^{12}~M_\odot$ in~\cite{2023arXiv230915986M}.

{We will show that a self-interacting cross section per particle mass of $\sigma/m\approx0.1~{\rm cm^2/g}$ provides an excellent fit to the measurement of JWST-ER1. This is the first test on SIDM from strong lensing objects at the highest redshift discovered to date. We also demonstrate that $\sigma/m\approx0.1~{\rm cm^2/g}$ can be consistent with strong lensing observations of early-type galaxies at $z\approx0.2$. These galaxies favor bottom-heavy Salpeter-like stellar initial mass functions~\citep{2010ApJ...709.1195T} and NFW-like dark matter halo profiles, see, e.g.,~\cite{2021MNRAS.503.2380S}. Our analysis is based on the lens model of J1636+4707 from~\cite{2021MNRAS.503.2380S}.} Additionally, we will discuss our SIDM constraints in the context of SIDM models that have been proposed to explain the diversity of dark matter distributions in other galactic systems.

\section{Initial conditions}
\label{sec:baryon}

In this section, we present initial conditions for modeling the mass distribution of the early-type galaxy JWST-ER1g.~\cite{2023NatAs.tmp....7V} used a S\'ersic profile~\citep{1963BAAA....6...41S} to fit the light distribution of stars in JWST-ER1g, and found that the effective radius is $R_e=1.9\pm0.2~{\rm kpc}$ and the S\'ersic index is $n=5.0\pm0.6$. The total stellar mass is $1.3^{+0.3}_{-0.4}\times10^{11}~M_\odot$, assuming a Chabrier initial mass function~\citep{2003PASP..115..763C}. 

In our study, we use a Hernquist profile to model the stellar distribution~\citep{1990ApJ...356..359H}
\begin{equation}
\label{eq:baryon}
\rho_b(r)= \frac{M_b }{ 2 \pi a^3} \frac{a^4}{r\left(r+a\right)^3},
\end{equation}
where $M_b$ is the total stellar mass and $a$ is the characteristic radius. We take $M_b = 1.3 \times 10^{11}~{M_{\odot}}$ and $a=0.79~{\rm kpc}$. Since the Hernquist profile is spherically symmetric and has a simple analytical form, it can be easily implemented in semi-analytical methods and N-body simulations that we will use.   

\begin{figure}[t!]
\centering
\includegraphics[scale=0.5]{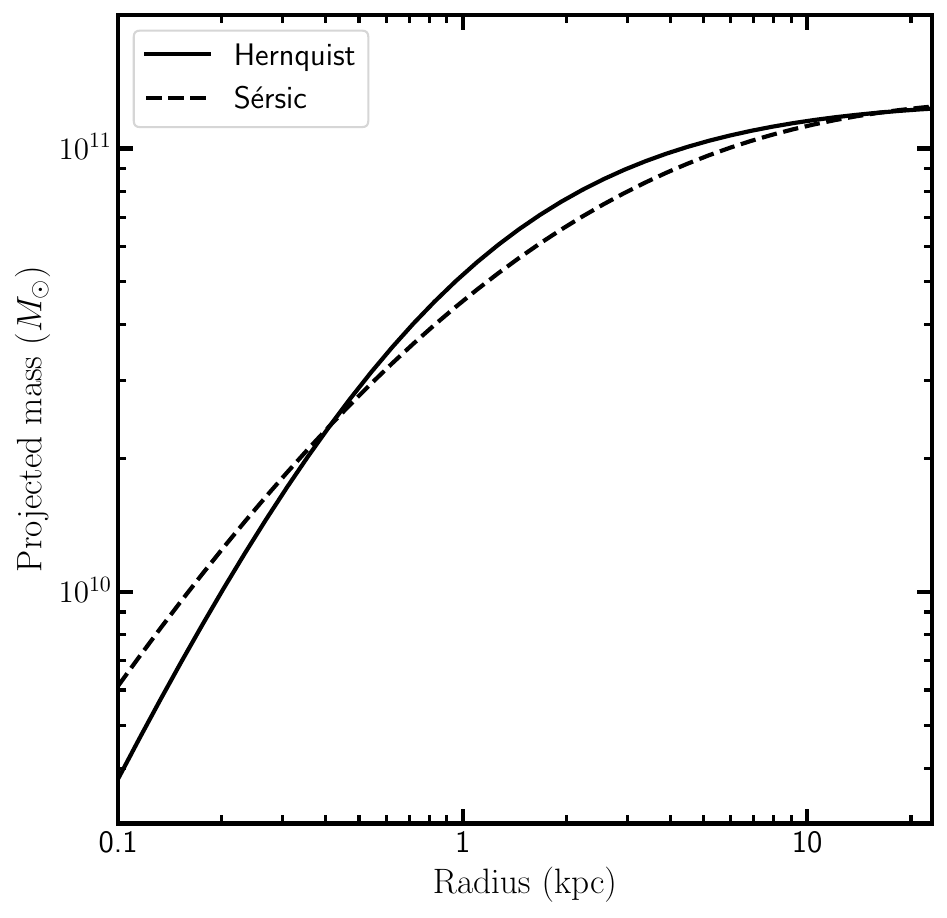}
\caption{The projected mass profile of the Hernquist profile used for modeling the stellar distribution of JWST-ERg, compared to the S\'ersic profile from~\cite{2023NatAs.tmp....7V}. }
\label{fig:masscomp}
\end{figure}

We take the profile in Equation~\ref{eq:baryon} to calculate the projected mass profile of the stars as
\begin{equation}
\label{eqn:projectedmass}
    M_b(R) = 2 \pi \int_0^{R} \int_{-\infty}^{+\infty} \rho_{b}(\sqrt{R'^2 + z^2})dz R'dR',
\end{equation}
and compare it to the S\'ersic profile with $R_e\approx1.9~{\rm kpc}$ and $n=5$ of JWST-ER1g~\citep{2023NatAs.tmp....7V}. Figure~\ref{fig:masscomp} shows the comparison between the Hernquist (solid-black) and S\'ersic (dashed-black) stellar profiles. The agreement is within $15\%$ for $R\gtrsim0.4~{\rm kpc}$. The Hernquist profile can underestimate the mass by $30\%$ towards the center $R\sim0.1~{\rm kpc}$, but even with the deficiency, the contraction effect is sufficiently strong.

For the dark matter halo of JWST-ER1g, we assume an NFW profile as the initial condition~\citep{Navarro_1997} ,
\begin{equation} 
\label{eq:nfw}
\rho_{\rm NFW}(r)=\frac{\rho_s r^3_s}{r(r+r_s)^2},
\end{equation}
where we take the scale density and radius as $\rho_{s} = 7.8 \times 10^{6} ~M_{\odot}/{\rm kpc^3}$ and $r_{s} = 70~\rm{kpc}$, respectively. The corresponding halo mass is $M_{200} = 3.0 \times 10^{13}~{M_{\odot}}$ and concentration $c_{200} = 4.6$ from the concentration-mass relation at $z=2$ with the scatter $0.11\mathrm{dex}$~\citep{2014MNRAS.441.3359D}. The concentration is $1\sigma$ above the cosmological median at $z\approx2$. The halo mass is consistent with the expectation from the stellar-halo mass relation~\citep{2013ApJ...770...57B}.~ {These halo parameters are chosen such that the contracted CDM halo can give rise to the mean value of the projected halo mass of JWST-ER1g within the Einstein radius, as we will show later.} Note we have used the following relations 
\begin{equation}
\rho_s=\frac{c^3_{200}}{3f(c_{200})}\Delta\rho_{\rm crit}(z),~r_s=\frac{r_{200}}{c_{200}},
\end{equation}
where $\Delta\approx200$, $\rho_{\rm crit}(z)$ is the critical density of the universe at redshift $z$, $r_{200}$ is the virial radius of the halo, and $f(c_{200})=\ln(1+c_{200})-c_{200}/(1+c_{200})$.

\section{CDM interpretation}
\label{sec:contra}

In response to the infall and condensation of baryons, a CDM halo becomes denser through the process of adiabatic contraction, and this effect can be modeled using a semi-analytical method~\citep{1986ApJ...301...27B}. In this work, we follow the one in~\cite{Gnedin_2004}, which was calibrated against hydrodynamical simulations. One introduces an adiabatic invariant $M(\bar{r}) r$, where $M(\bar{r})$ is the total mass within the orbital-averaged radial position $\bar{r}$. It follows the relation $\bar{x}=A x^w$ with $x \equiv r / r_{200}$, $A \approx 0.85 \pm 0.05$ and $w \approx 0.8 \pm 0.02$. Assuming that baryonic and dark matter follow the same initial radial profile, we can use the adiabatic invariant and obtain the final radius of dark matter particles $r_f$, which are initially located at $r$,
\begin{equation}
\label{eq:ad}
\frac{r}{r_f}=1-f_b+\frac{M_b(\bar{r}_f)}{M_i(\bar{r})},
\end{equation}
where $f_b\equiv M_b/M_{200}$ is the baryon mass fraction, $M_b(\bar{r}_f)$ is the final baryon mass profile, and $M_i(\bar{r})$ the initial total mass profile. In our study, the final baryon and initial total density profiles are given by Equations~\ref{eq:baryon} and~\ref{eq:nfw}, respectively. For given $r$, we solve Equation~\ref{eq:ad} to obtain $r_f$, and obtain the density profile of a contracted CDM halo as $M_{{\rm dm}}(r_f)=(1-f_b)M_{i}(r)$. In practice, we use a public code provided by~\cite{2023MNRAS.521.4630J} to calculate $M_{\rm dm} (r_f)$.

\begin{figure}[t!]
\centering
\includegraphics[scale=0.5]{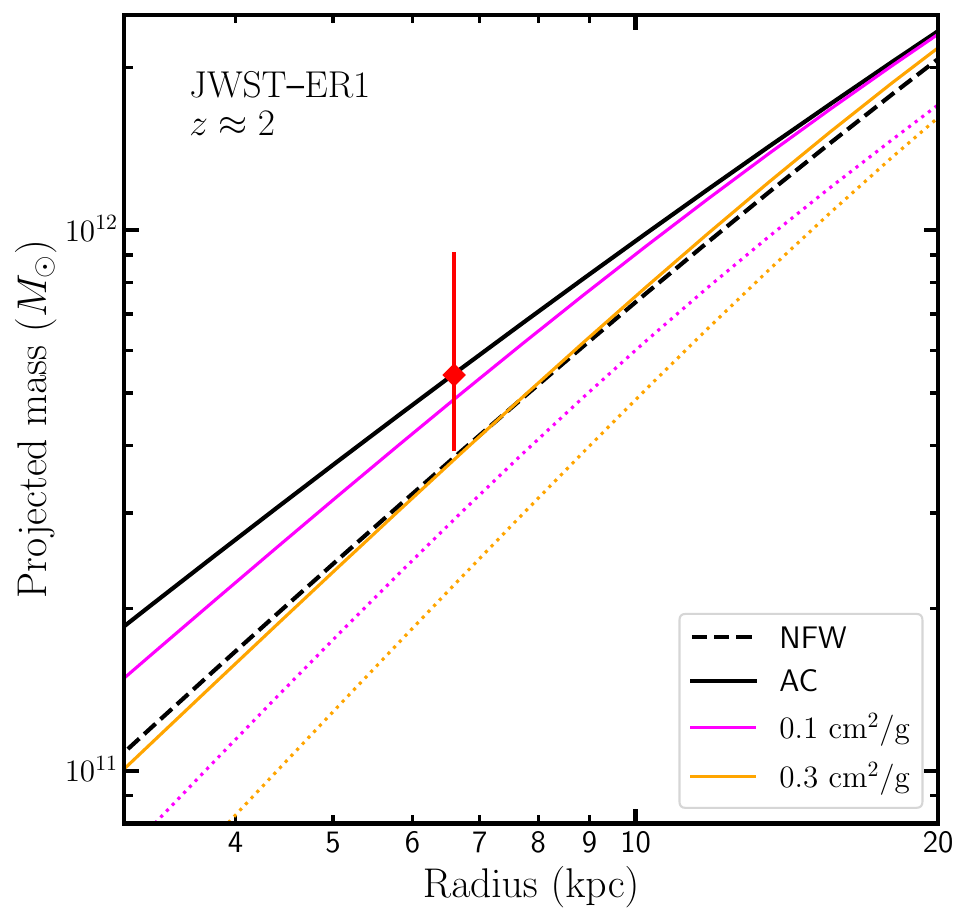}
\caption{JWST-ER1g: The projected halo mass profile of the contracted CDM halo (solid-black), the initial NFW halo (dashed-black), the SIDM halos with $\sigma/m=0.1~{\rm cm^2/g}$ (solid-magenta) and $0.3~{\rm cm^2/g}$ (solid-orange). The corresponding SIDM halos without including the influence of baryons are shown for comparison (dotted). The measured value is $5.4^{+3.7}_{-1.5}\times10^{11}~{M_\odot}$ within the Einstein radius $6.6~{\rm kpc}$ (red diamond), after subtracting the baryon mass from the total enclosed mass~\citep{2023NatAs.tmp....7V}.}
\label{fig:jwst}
\end{figure}

Figure~\ref{fig:jwst} shows the projected mass profile of the contracted (solid-black) and initial (dashed-black) CDM halos. We see that the contracted CDM halo is sufficiently dense to be consistent with the projected halo mass within the Einstein radius $6.6~{\rm kpc}$ of JWST-ER1g $5.4^{+3.7}_{-1.5}\times10^{11}~M_\odot$ (red diamond), while the initial halo is too shallow. The projected mass within $6.6~{\rm kpc}$ is increased by $30\%$ from $3.8\times10^{11}~M_\odot$ to $5.4\times10^{11}~M_\odot$ due to adiabatic contraction. Thus the CDM scenario can explain the strong lensing observations of JWST-ER1 even with a bottom-light Chabrier initial mass function.

The choice of $M_{200}$ and $c_{200}$ values is not unique, but we have checked that the halo mass is the dominant factor. We have checked that the halo mass is the dominant factor in the fit. If we take $M_{200}=10^{13}~{M_\odot}$, a factor of $3$ smaller than the fiducial value we assume, the concentration needs to be increased to $c_{200}\approx9.5$, which is $4\sigma$ above the cosmological median at $z\approx2$~\citep{2014MNRAS.441.3359D}. For a fixed halo mass, the projected mass of a contracted CDM halo has a mild dependence on the initial concentration; see Appendix~\ref{app:app1} for details. 

From Figure~\ref{fig:jwst}, we also see that the impact of adiabatic contraction on the halo is beyond the central region characterized by the stellar effective radius $1.9~{\rm kpc}$. This is because the initial distribution of gas is spread out over the entire halo, and as the gas cools and condenses towards the center, the entire halo contracts~\citep{1986ApJ...301...27B,Gnedin_2004}. We have further checked that the contraction effect is strong enough to explain the high mass of JWST-ER1g even if the stellar potential grows from the center; see Appendix~\ref{app:app2} for the demonstration using N-body simulations. 

In Appendix~\ref{app:app3}, we present a similar analysis using the measurements from~\cite{2023arXiv230915986M}. After the contraction effect is included, a CDM halo with $M_{200}\approx10^{13}~M_\odot$ is too dense to be consistent with the measured lens mass in~\cite{2023arXiv230915986M} and a $6\times10^{12}~{M_\odot}$ halo with low concentration is favored, which is on the lower end of the stellar mass-halo mass relation for the given stellar mass~\citep{2013ApJ...770...57B}.

\section{SIDM interpretation}
\label{sec:sidmframe}

In this section, we study the interpretation of JWST-ER1 in the SIDM scenario~\citep{Tulin_2018,Adhikari:2022sbh}. Dark matter self-interactions could thermalize the inner halo and produce a shallow density core, which seemed contradictory to the observations of JWST-ER1. However, in the presence of baryons, SIDM thermalization actually leads to a small core and a high inner halo density~\citep{PhysRevLett.113.021302,Kamada:2016euw,Creasey:2017qxc,Elbert:2016dbb,Robertson:2017mgj,Robertson:2018anx,Robertson:2020pxj,Sameie:2018chj,Despali:2018zpw,Santos-Santos:2019vrw,Feng:2020kxv,Sameie:2021ang,Zhong:2023yzk}. 

To model the SIDM density profile for JWST-ER1g, we follow a semi-analytical method in~\cite{PhysRevLett.113.021302,PhysRevLett.116.041302}, which has been tested extensively using hydrodynamical SIDM simulations, see, e.g.,~\cite{Robertson:2017mgj,Robertson:2020pxj}. In the inner region, where SIDM thermalization occurs over the age of the galaxy, the halo density profile $\rho_{\rm iso}(r)$ can be obtained by solving the isothermal Jeans equation
\begin{equation}
\label{eq:iso}
   \sigma^2_0\nabla^2\ln\rho_{\rm iso}=-4\pi G (\rho_{\rm iso}+\rho_{b}),
\end{equation}
where $G$ is the Newton constant, $\sigma_0$ is the 1D velocity dispersion of dark matter particles, $\rho_{b}(r)$ is the final baryon density profile. In the outer region, where the thermalization effect can be negligible, the halo is in the collisionless limit and can be modeled with a CDM density profile of $\rho_{\rm cdm}(r)$. The boundary is set by the characteristic radius $r_1$, at which self-scattering occurs once per age of the galaxy $t_{\rm age}$, i.e.,  
\begin{equation}
\label{eq:r1}
 \frac{\sigma}{m} \rho_{\rm iso}(r_1)  \frac{4 }{\sqrt{\pi}}\sigma_0 t_{\rm{age}} = 1,
\end{equation}
where $\sigma/m$ is the self-interacting cross section per particle mass. The full SIDM density profile is composed of two parts: $\rho_{\rm iso}$ for $r<r_1$ and $\rho_{\rm cdm}$ for $r>r_1$; they satisfy the matching conditions $\rho_{\rm iso}(r_1)\approx\rho_{\rm cdm}(r_1)$ and $M_{\rm iso}(r_1)\approx M_{\rm cdm}(r_1)$. 

We set $\rho_{b}(r)$ to be the Hernquist profile in Equation~\ref{eq:baryon} and $\rho_{\rm cdm}(r)$ to be the contracted CDM profile shown in Figure~\ref{fig:jwst} (solid-black), and $t_{\rm age}=3.4~{\rm Gyr}$, corresponding to the redshift $z\approx2$. We also assume a constant cross section, which should be regarded an effective cross section for the halo~\citep{Yang_2022,Yang:2023jwn,Outmezguine:2022bhq,Yang:2022zkd,Fischer:2023lvl}. For a given value of $\sigma/m$, we solve Equations~\ref{eq:iso} and~\ref{eq:r1} to find the SIDM density profile. In our numerical study, we again use the public code by~\cite{2023MNRAS.521.4630J} to obtain the SIDM density profile, and further test it using N-body simulations; see Appendix~\ref{app:app2}. 

Figure~\ref{fig:jwst} shows the projected halo mass profile of the SIDM halos assuming $\sigma/m=0.1~{\rm cm^2/g}$ (solid-magenta) and $0.3~{\rm cm^2/g}$ (solid-orange). The projected halo mass decreases as the cross section increases. At $z\approx2$, the halo is at the stage of core formation and a larger cross section leads to a larger core and smaller inner density. {With the measurement uncertainty, the strong lensing object JWST-ER1 favors a small cross section of $\sigma/m\lesssim0.3~{\rm cm^2/g}$ in a $10^{13}~{M_\odot}$ halo, and $\sigma/m\approx0.1~{\rm cm^2/g}$ can provide an excellent fit. In Appendix~\ref{app:app3}, we also show that the SIDM constraints are slightly stronger when using the measurements in~\cite{2023arXiv230915986M}, but overall, they are largely robust to the exact value of the lens mass of JWST-ER1. We have further confirmed it using a Bayesian analysis by varying $\sigma/m$, $M_{200}$, and $c_{200}$; see Appendix~\ref{app:mcmc}.}

Figure~\ref{fig:jwst} also shows the projected mass profile for the SIDM halos with $\sigma/m=0.1~{\rm cm^2/g}$ (dotted-magenta) and $0.3~{\rm cm^2/g}$ (dotted-orange) {\it without} including the baryons, i.e., neglecting the term $\rho_b(r)$ in Equation~\ref{eq:iso}. These SIDM halos are too shallow to explain the lens mass of JWST-ER1. In fact, they have a $20~{\rm kpc}$ density core. In contrast, the SIDM thermalization leads to the increase of the inner density by several orders of magnitude in the presence of the baryons. Thus JWST-ER1 provides a crucial test on the mechanism of tying dark matter and baryon distributions in SIDM, which was first proposed in~\cite{PhysRevLett.113.021302}. 

It is interesting to note that for $\sigma/m=0.3~{\rm cm^2/g}$, the SIDM density profile at $t_{\rm age}\approx3.4~{\rm Gyr}$ well resembles the initial NFW profile {\it without} adiabatic contraction. From Equation~\ref{eq:r1}, we see that $\sigma/m$ and $t_{\rm age}$ are degenerate in determining the SIDM density profile. In other words, the density profile with $\sigma/m=0.3~{\rm cm^2/g}$ and $t_{\rm age}\approx3.4~{\rm Gyr}$ is equivalent to that with $\sigma/m=0.1~{\rm cm^2/g}$ and $t_{\rm age}\approx10~{\rm Gyr}$. Thus in SIDM with $\sigma/m\approx0.1~{\rm cm^2/g}$ dark matter halos of early-type galaxies at $z\sim0.2$ are expected to be NFW-like, i.e., their density profiles are shallower than those predicted in contracted CDM halos, as we demonstrate in the next section.

\section{Early-type galaxies at $z\sim0.2$}
\label{sec:evo}

\begin{figure}[t!]
\centering
\includegraphics[scale=0.5]{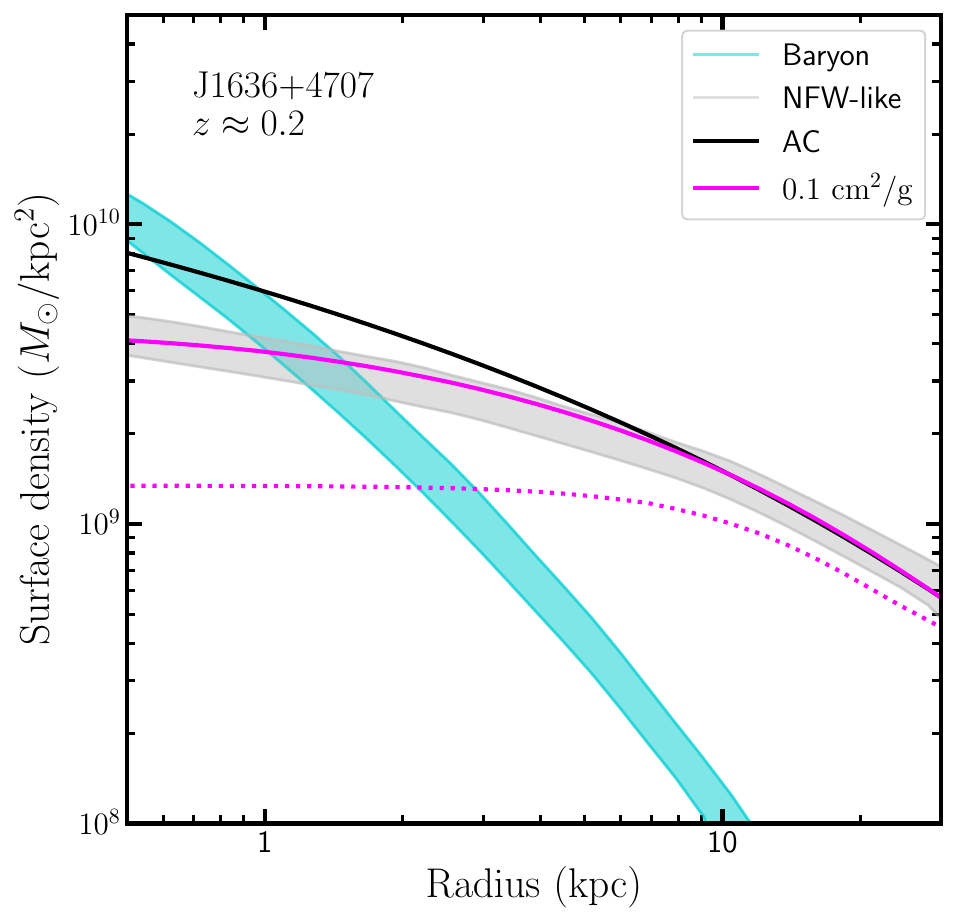}
\caption{J1636+4707: The surface density profile of the SIDM halo with $\sigma/m=0.1~{\rm cm^2/g}$ (solid-magenta), the stars (cyan) and the NFW-like halo (gray) from observations ($1\sigma$ band)~\citep{2021MNRAS.503.2380S}. For comparison, the CDM halo with adiabatic contraction (solid-black) and the SIDM halo without baryons (dotted-magenta) are shown.}
\label{fig:objectJ1636}
\end{figure}

\cite{2021MNRAS.503.2380S} analyzed imaging data of a sample of $23$ galaxy-galaxy lenses, and found that the dark matter halos of the elliptical galaxies at $z\sim0.2$ are NFW-like and on average there are no signs of adiabatic contraction. This result is robust to the assumption of stellar mass-to-light ratio models. {With the insights gained in the last section, we explore the viability of the SIDM interpretation for early-type galaxies at $z\sim0.2$.}

{As a demonstration, we take the lens mass model of J1636+4707 from~\cite{2021MNRAS.503.2380S}}. We use the Hernquist profile in Equation~\ref{eq:baryon} to fit the observed stellar density profile in~\citet{2021MNRAS.503.2380S} and determine the relevant parameters as $M_b=2.2\times10^{11}~M_\odot$ and $a=2.7~{\rm kpc}$. Note that J1636+4707 has a stellar size of $R_e\approx6.5~{\rm kpc}$, much larger than $R_e=1.9~{\rm kpc}$ of JWST-ER1g. The initial NFW halo parameters are found to be $\rho_s=5.6\times10^6~M_\odot/{\rm kpc^3}$ and $r_s=67~{\rm kpc}$, corresponding to a halo mass of $M_{200}=3.0\times10^{13}~M_\odot$ at $z\approx0.2$. We set $\sigma/m=0.1~{\rm cm^2/g}$ and $t_{\rm age}=11.3~{\rm Gyr}$ ($z\approx0.2$). 

Figure~\ref{fig:objectJ1636} shows the surface density profile of the SIDM halo (solid-magenta), as well as the NFW-like halo (gray) and the baryon component (cyan) inferred from observations within the $1\sigma$ credible range~\citep{2021MNRAS.503.2380S}. For comparison, we also show the CDM halo with adiabatic contraction (solid-black) and the SIDM halo without baryons (dotted-magenta). The SIDM density profile is well consistent with the NFW-like profile from the lensing measurements~\citep{2021MNRAS.503.2380S}. On the other hand, the contracted CDM halo is too dense towards the inner regions, while the SIDM halo without baryons has a $20~{\rm kpc}$ core and is too shallow. We have further confirmed these results using controlled CDM and SIDM N-body simulations; see Appendix~\ref{app:app2}. 

{More work is needed to further test the SIDM interpretation of early-type galaxies at $z\approx0.2$. In fact, the galaxies in~\cite{2021MNRAS.503.2380S} exhibit rather diverse stellar distributions, and it would be important to check if $\sigma/m\approx0.1~{\rm cm^2/g}$ can fit to all of them. Additionally, in contrast to a Chabrier initial mass function for JWST-ER1g~\citep{2023NatAs.tmp....7V},  the galaxies in~\cite{2021MNRAS.503.2380S} systemically favor a bottom-heavy, Salpeter form~\citep{2010ApJ...709.1195T}, thus a full explanation of their NFW-like halo properties must be related to the stellar initial mass function as well. We will leave these exciting topics for future work.}

\section{Discussion and Conclusion}
\label{sec:conclusion}

We have shown that a contracted CDM halo can explain the high halo density of JWST-ER1g at $z\approx2$. {This in turn puts constraints on the dark matter self-interaction cross section in $10^{13}~M_\odot$ halos: a small cross section of $\sigma/m\lesssim0.3~{\rm cm^2/g}$ is generally favored, and $\sigma/m\approx0.1~{\rm cm^2/g}$ gives rise to an excellent fit.} These constraints are largely robust to the exact value of the lens mass of JWST-ER1. Intriguingly, SIDM with $\sigma/m\approx0.1~{\rm cm^2/g}$ may also explain NFW-like halo properties of early-type galaxies at low redshift $z\approx0.2$, as the SIDM halo further evolves and the density profile becomes shallower. 

\cite{Despali:2018zpw} performed hydrodynamical SIDM simulations of formation of early-type galaxies assuming $\sigma/m=1~{\rm cm^2/g}$. They showed that when baryons are included, the difference between CDM and SIDM predictions becomes small and they are largely consistent with various observational constraints; see~\cite{Despali:2022vgq,Mastromarino:2022hwx,McDaniel:2021kpq} for further investigations. 

Overall, their findings are in good agreement with ours, i.e., the baryons play a vital role in explaining the observations of early-type galaxies. We have checked that the simulated SIDM halos in~\cite{Despali:2018zpw} that could potentially host JWST-ER1g are not dense enough; see their Figure 7. Nevertheless, their most massive SIDM halos are cored at $z\approx0.2$, shallower than the corresponding CDM counterparts, in alignment with what we found. More work is needed to test SIDM predictions of early-type galaxies for $\sigma/m\approx0.1~{\rm cm^2/g}$ using hydrodynamical simulations. 

Studies show that $\sigma/m\lesssim 0.1~{\rm cm^2/g}$ in $10^{15}~M_\odot$ cluster halos~\citep{PhysRevLett.116.041302,10.1093/mnras/stab3241,2022A&A...666A..41E,Adhikari:2024aff}, and $\sigma/m\lesssim1~{\rm cm^2/g}$ in $10^{14}~{M_\odot}$ group halos~\citep{Sagunski_2021}. Our constraints are consistent with those previous studies, but on a lower mass scale $\sim10^{13}~M_\odot$. On the other hand, for dwarf galaxies, $\sigma/m\gtrsim10~{\rm cm^2/g}$ is favored to explain the diversity in the dark matter distribution, see, e.g.,~\cite{Valli:2017ktb,Ren:2018jpt,Sameie:2019zfo,Zavala:2019sjk,Kahlhoefer:2019oyt,Correa:2020qam,Ray:2022ydr,Silverman:2022bhs,Correa:2022dey,Yang:2022mxl,Nesti:2023tid,Nadler:2023nrd,Zhang:2024ggu}. For a viable SIDM model, the cross section must be velocity-dependent, decreasing towards massive halos~\citep{Tulin_2018}. Interestingly, in the velocity-dependent SIDM model proposed in~\cite{Nadler:2023nrd} to explain observations of extreme halo diversity, the predicted effective cross section is $0.1~{\rm cm^2/g}$ in $10^{13}~{M_\odot}$ halos, consistent with our constraints from the early-type galaxies.

In the future, more objects like JWST-ER1 could be discovered with JWST and they would further test our CDM and SIDM interpretations. It would be of great interest to perform SIDM fits to a large sample of early-type galaxies at low redshift, such as those in~\cite{2021MNRAS.503.2380S,2023arXiv231109307T}. Cosmological hydrodynamical simulations would be needed to further understand the formation and evolution of early-type galaxies in both CDM and SIDM scenarios.

\begin{acknowledgments}
This work was supported by the John Templeton Foundation under grant ID
\#61884 and the U.S. Department of Energy under grant No. de-sc0008541. The opinions expressed in this publication are those of the authors and do not necessarily reflect the views of the John Templeton Foundation. 
\end{acknowledgments}

\vspace{2mm}

\software{NumPy \citep{harris2020array},  
          SciPy \citep{2020SciPy-NMeth}, 
          Matplotlib \citep{Hunter:2007},
          Colossus \citep{2018ApJS..239...35D},
          emcee \citep{2013PASP..125..306F},
          corner \citep{corner}
          }

\appendix

\section{Varying model parameters}
\label{app:app1}

\begin{figure*}[ht]
\centering
\includegraphics[scale=0.4]{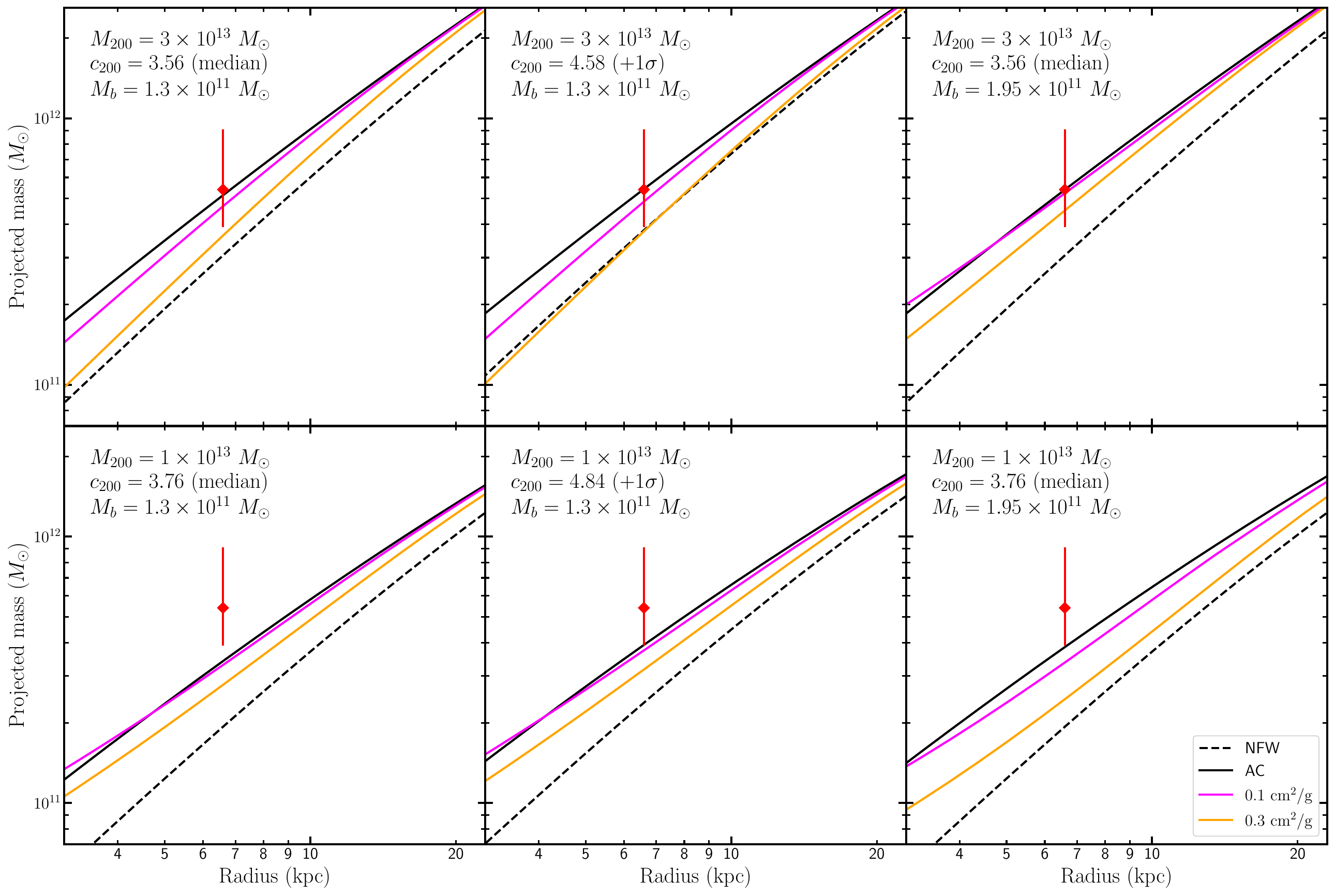}
\caption{The projected halo mass profiles for six different sets of the model parameters for JWST-ER1g: $M_{200}$, $c_{200}$, and $M_b$. The line style is the same as in Figure~\ref{fig:jwst}.}
\label{fig:inthalocomp}
\end{figure*}

Figure~\ref{fig:inthalocomp} shows the projected halo mass profiles for six different sets of the model parameters in the case of JWST-ER1g. For both CDM and SIDM halos, the projected halo mass profiles have a mild dependence on the initial halo concentration and the total baryon mass. If the halo mass is $10^{13}~{M_\odot}$ or lower, the projected mass within the Einstein radius will be too low to explain the high dark matter density of JWST-ER1g, even if we increase the total baryon mass to $2.0\times10^{11}~M_\odot$, $2\sigma$ higher than the mean reported in~\cite{2023NatAs.tmp....7V}.

\section{Controlled N-body simulations}
\label{app:app2}

\begin{figure*}[ht]
\centering
\includegraphics[scale=0.5]{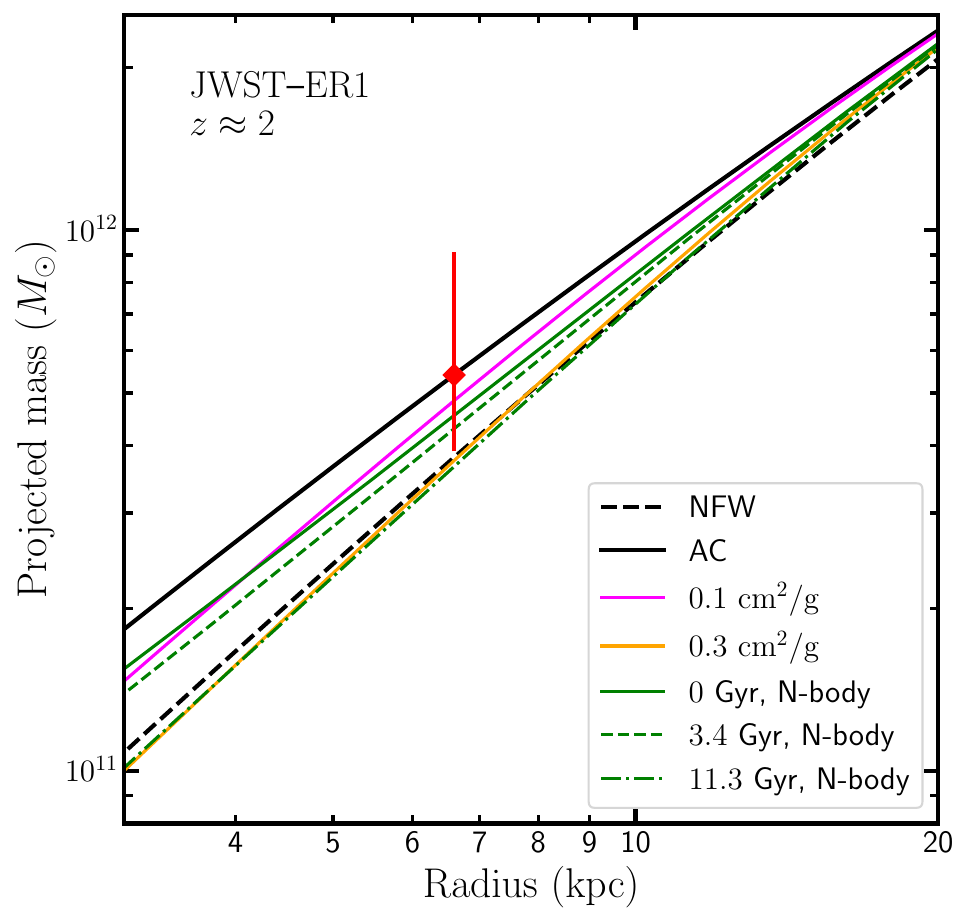}~~~~~
\includegraphics[scale=0.5]{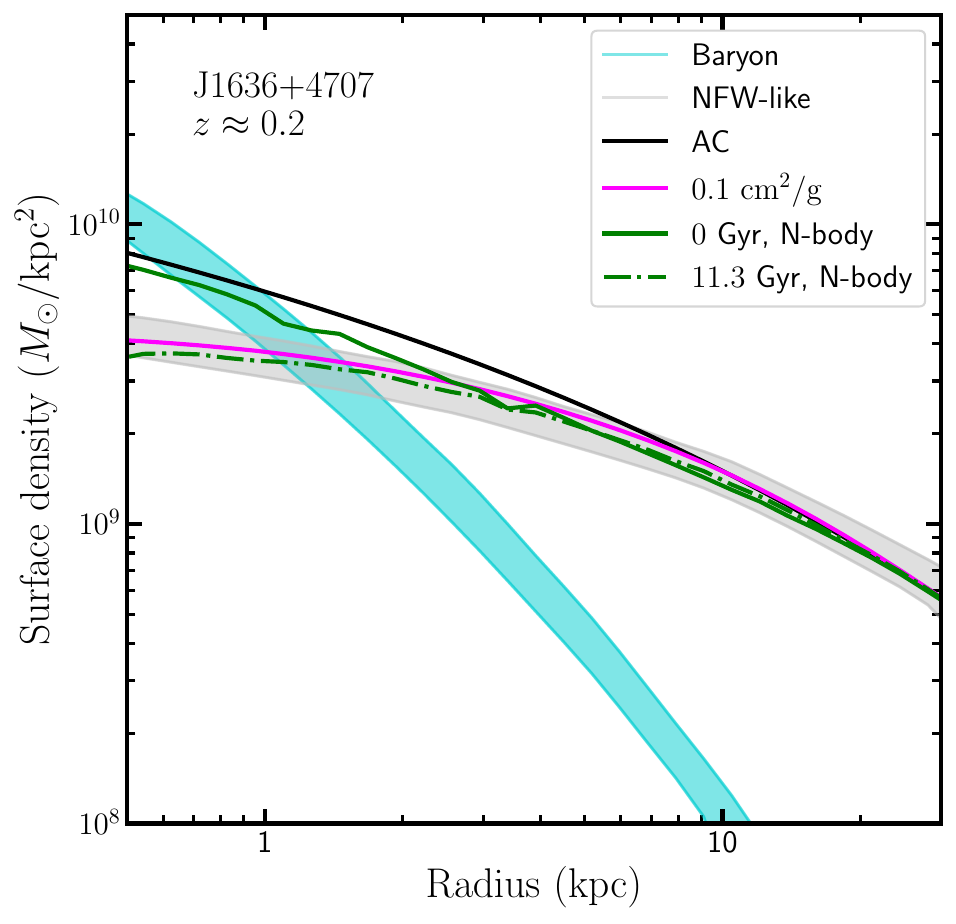}
\caption{Left panel: The projected halo mass profile of the simulated SIDM halo at $t=0~{\rm Gyr}$ (solid-green), $3.4~{\rm Gyr}$ (dashed-green), and $11.3~{\rm Gyr}$ (dashed-dotted-green) for JWST-ER1g. Right panel: The surface density profile of the simulated SIDM halo at $t=0~{\rm Gyr}$ (solid-green) and $11.3~{\rm Gyr}$ (dashed-green) for J1636+4707. For both simulated halos, $\sigma/m=0.1~{\rm cm^2/g}$. Other curves are the same as those in Figures~\ref{fig:jwst} and~\ref{fig:objectJ1636}.}
\label{fig:sim}
\end{figure*}

We perform controlled N-body simulations to model the CDM and SIDM halos of JWST-ER1g and J1636+4707. For JWST-ER1g, we take the initial NFW halo parameters as $\rho_s=7.8\times10^6~{M_\odot}/{\rm kpc^3}$ and $r_s=70~{\rm kpc}$. At the center of the halo, we grow a baryonic potential following a Hernquist profile. The scale radius is fixed to $a=0.79~{\rm kpc}$ initially and remains a constant, while the mass grows linearly to $M_b=1.3\times10^{11}~M_\odot$ within $5~{\rm Gyr}$. Similarly, for J1636+4707, the initial halo parameters are $\rho_s=5.6\times10^6~{M_\odot}/{\rm kpc^3}$ and $r_s=67~{\rm kpc}$, and a growing Hernquist potential with $a=2.7~{\rm kpc}$ that reaches $M_b=2.2\times10^{11}~M_\odot$. For both cases, we let their CDM halos fully relax in the presence of the baryonic potential and obtain their contracted CDM density profiles. Then, we turn on dark matter self-interactions with $\sigma/m=0.1~{\rm cm^2/g}$ and follow gravothermal evolution of the halos. We use the public code~\texttt {SpherIC}~\citep{Garrison-Kimmel:2013yys} to generate initial conditions and the code~\texttt {GADGET-2}~\citep{Springel:2000yr-GADGET2,Springel:2005mi-GADGET2}, implemented with an SIDM module~\citep{Yang_2022}, to perform N-body simulations. The total number of simulation particles is $10^{6}$ ($2\times10^6$) for JWST-ER1g (J1636+4707), the particle mass is $3.9\times10^7~M_\odot$ ($1.9\times10^7~M_\odot$). For both cases, the softening length is set to $1~{\rm kpc}$. Note when reporting evolution time of the simulated SIDM halos, we do not account for the time for growing the potential and relaxing the CDM halos. 

Figure~\ref{fig:sim} (left panel) shows the projected mass profile of the contracted CDM halo ($t=0~{\rm Gyr}$, solid-green), the SIDM halo at $t=3.4~{\rm Gyr}$ (dashed-green) and $11.3~{\rm Gyr}$ (dashed-dotted-green) from our N-body simulations, in the case of JWST-ER1g. Compared to the contracted CDM halo induced by baryon infall and condensation (solid-black), our simulated CDM halo is less dense and the baryons have less impact over large radii. This is because we purposely grow the potential from the center of the halo, while the semi-analytical method in~\cite{Gnedin_2004} is calibrated against cosmological hydrodynamical simulations that take into account the contraction effect on the entire halo. Even with the conservative approach to model adiabatic contraction, our contracted CDM halo is sufficiently dense to be consistent with the measurement within the uncertainty (red diamond), further confirming our result that a CDM halo with adiabatic contraction can explain the high halo density of JWST-ER1g.   

Furthermore, the projected mass of the simulated SIDM halo at $t=3.4~{\rm Gyr}$ (dashed-green) agrees with that of the SIDM halo with $\sigma/m=0.1~{\rm cm^2/g}$ from the semi-analytical method (solid-magenta) within $10\%$. The latter is slightly denser, because it was produced by matching to the contracted CDM halo based on~\cite{Gnedin_2004} (solid-black); see Equation~\ref{eq:r1} and we have set $t_{\rm age}=3.4~{\rm Gyr}$.  We also show the projected mass profile of the simulated SIDM halo at $t=11.3~{\rm Gyr}$, corresponding to $z\approx0.2$, and it is close to the initial NFW profile (dashed-black). Thus SIDM predicts that the halo of JWST-ER1 would be NFW-like if it further evolved from $z\approx2$ to $0.2$. It also well agrees with the SIDM halo with $\sigma/m=0.3~{\rm cm^2/g}$ at $t_{\rm age}=3.4~{\rm Gyr}$ from the semi-analytical method, confirming our expectation from the scaling relation $t_{\rm age}\propto(\sigma/m)^{-1}$ as indicated in Equation~\ref{eq:r1}. 
 
Figure~\ref{fig:sim} (right panel) shows the surface density of the SIDM halo at $t=11.3~{\rm Gyr}$ (dashed-dotted-green), as well as the contracted CDM halo ($t=0~{\rm Gyr}$, solid-green) from our controlled N-body simulations, in the case of J1636+4707. We again see that the simulated SIDM halo becomes NFW-like after gravothermal evolution, and its surface density is well within the range from the measurement~\citet{2021MNRAS.503.2380S} (gray band). Since the baryonic potential grows from the center of the halo in our N-body simulations, its impact on the halo has a smaller range compared to that using the method in~\cite{Gnedin_2004} (solid-black).

\section{Alternative fits}
\label{app:app3}

\begin{figure*}[ht]
\centering
\includegraphics[scale=0.5]{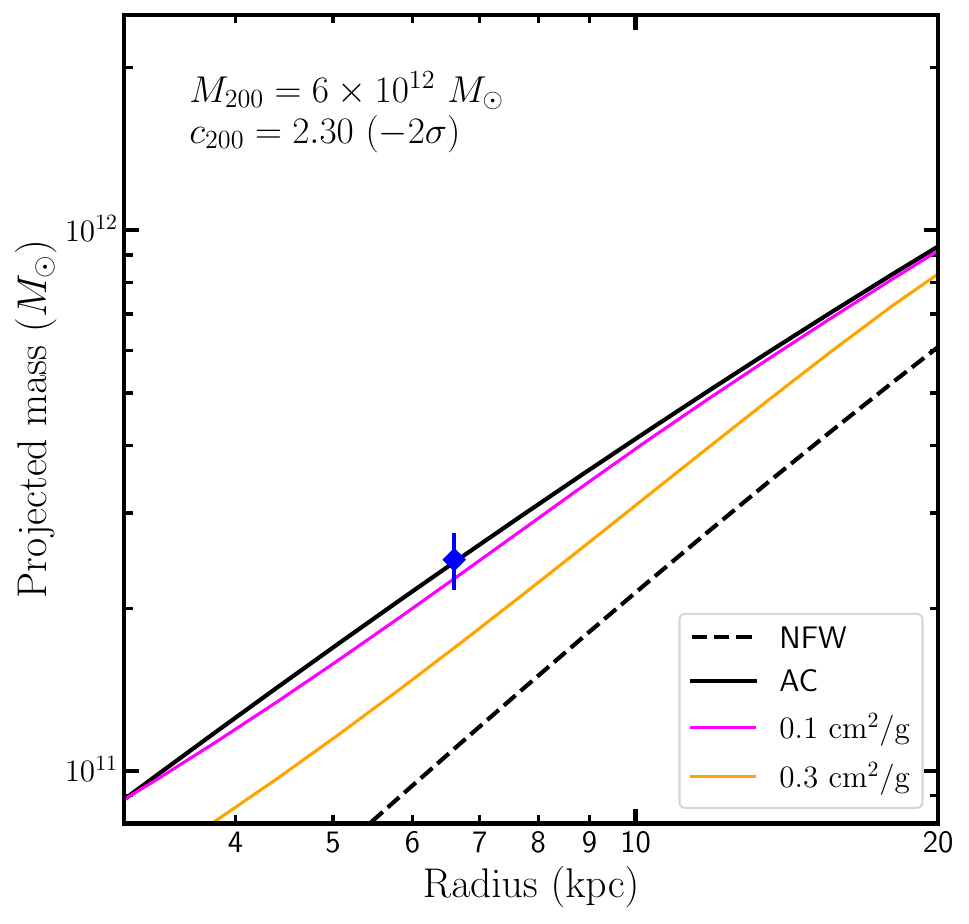}~~~~~
\includegraphics[scale=0.5]{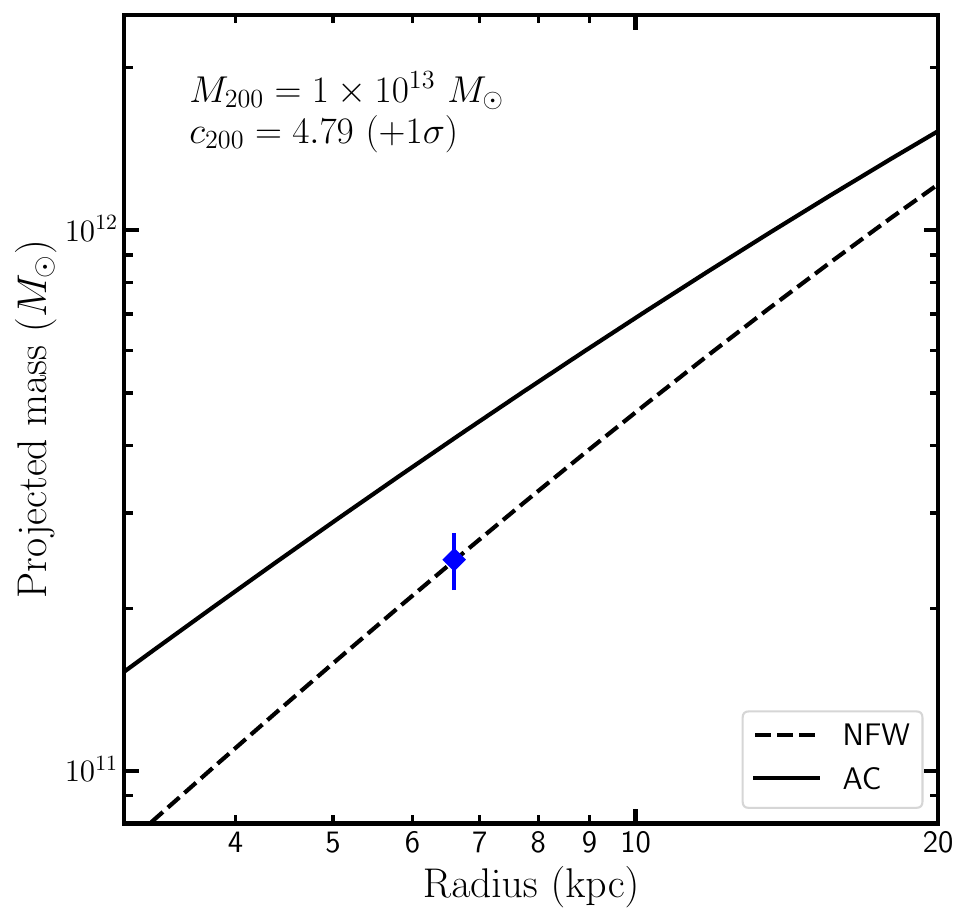}
\caption{Left panel: The projected mass profile of the contracted CDM halo (solid-black), the initial NFW halo
(dashed-black), the SIDM halos with $\sigma/m=0.1~{\rm cm^2/g}$ (solid-magenta) and $0.3~{\rm cm^2/g}$ (solid-orange) for JWST-ER1g based on the measurement in~\cite{2023arXiv230915986M} (blue diamond). Right panel: The projected mass profile of the contracted CDM halo (solid-black) and the initial NFW halo (dashed-black), where the halo parameters are chosen such that the NFW halo {\it without} adiabatic contraction could explain the measurement in~\cite{2023arXiv230915986M} (blue diamond). }
\label{fig:jwst2}
\end{figure*}

We perform a fit to the projected halo mass of JWST-ER1g reported in~\cite{2023arXiv230915986M}, i.e., $(2.46\pm0.30)\times10^{11}~M_\odot$. We use the Hernquist profile to model the stellar distribution and fix the parameters as $M_b=1.4\times10^{11}~M_\odot$ and $a=0.62~{\rm kpc}$. The halo mass is $M_{200}=6.0\times10^{12}~M_\odot$ and concentration is $c_{200}=2.30$, which is $2\sigma$ below the median at $z\approx2$. The corresponding initial NFW parameters are $\rho_s=1.9\times10^6~M_\odot/{\rm kpc^3}$ and $r_s=80~{\rm kpc}$. 

Figure~\ref{fig:jwst2} (left panel) shows the projected halo mass profile of the contracted CDM halo (solid-black), the initial NFW halo (dashed-black), as well as the SIDM halos with $\sigma/m=0.1~{\rm cm^2/g}$ (solid-magenta) and $0.3~{\rm cm^2/g}$ (solid-orange). The contracted halo is sufficiently dense to be consistent with the measurement in~\cite{2023arXiv230915986M} (blue diamond). In contrast, the NFW halo is too shallow. The constraints on SIDM are slightly stronger than those based on~\cite{2023NatAs.tmp....7V}, as~\cite{2023arXiv230915986M} reported smaller uncertainties in the measurement of the projected halo mass. Overall, our SIDM constraints from the JWST-ER1 object are largely robust to the exact value of the lens mass, and it is clear that $\sigma/m\approx0.1~{\rm cm^2/g}$ is allowed. 

Additionally, as shown in Figure~\ref{fig:jwst2} (right panel), a $10^{13}~M_\odot$ NFW halo could be dense enough to explain the measurement in~\cite{2023arXiv230915986M}. However, the contracted CDM halo is too dense. 

\section{The MCMC fit}
\label{app:mcmc}

\begin{figure*}
\centering
\includegraphics[scale=0.7]{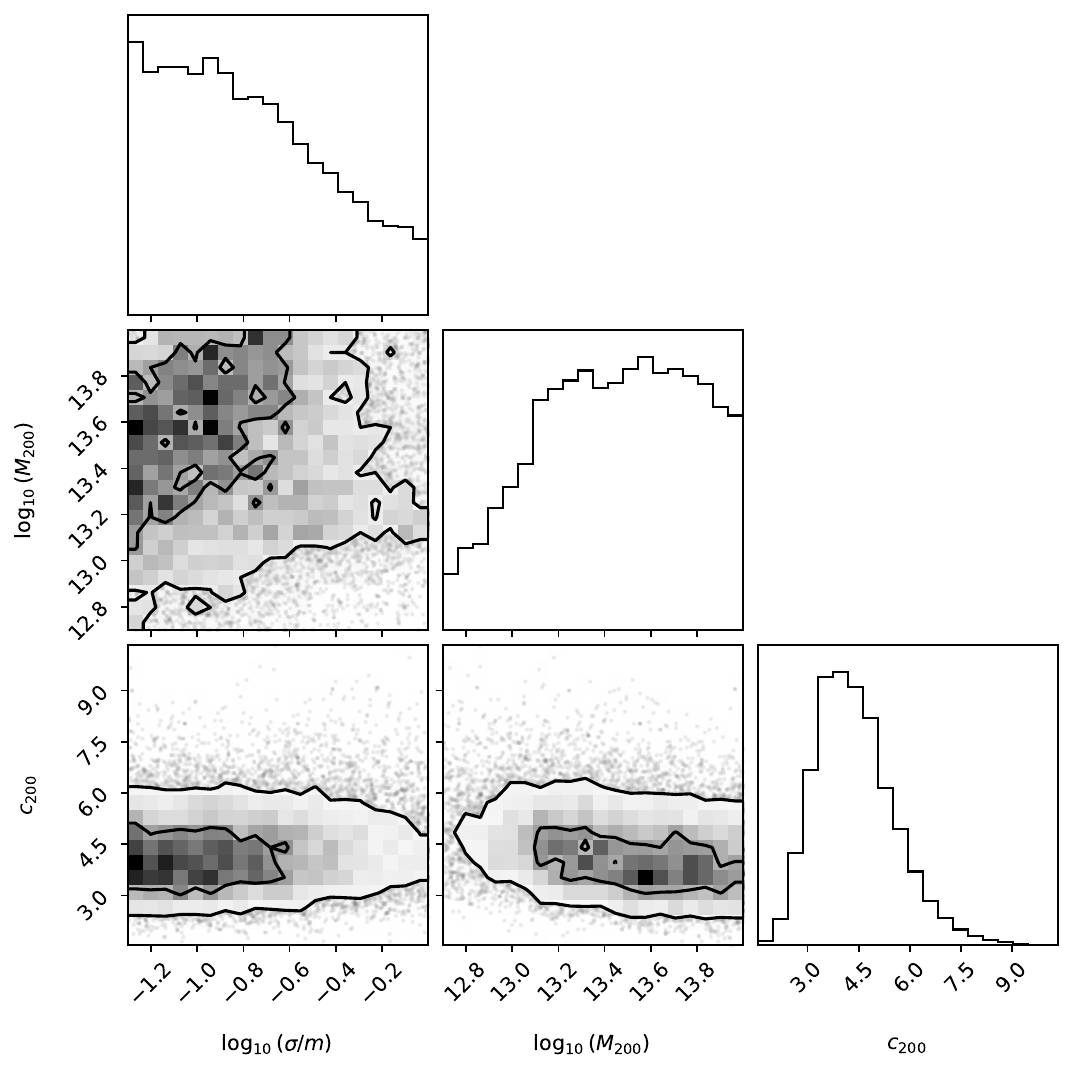}
\caption{The posterior distribution of the cross section $\sigma/m$, halo mass $M_{200}$, and concentration $c_{200}$ for JWST-ER1g with the measurement from \citet{2023NatAs.tmp....7V}, where the contours denote $1\sigma$ and $2\sigma$ confidence regions.}
\label{fig:corner}
\end{figure*}

{To further explore possible correlations between the cross section and halo mass, we perform a Bayesian analysis for JWST-ER1g based on the measurement in~\citet{2023NatAs.tmp....7V}, assuming the Chabrier initial stellar mass function. We utilize the MCMC method to sample the SIDM and halo parameters: a flat prior in a range of $\log_{10}(0.05)\leq\log_{10}(\frac{\sigma/m}{{\rm cm^2/g}})\leq\log_{10}(1)$ for the cross section, a flat prior in a range of $\log_{10}(5\times10^{12})\leq\log_{10}(\frac{M_{200}}{M_\odot})\leq\log_{10}(10^{14})$ for the halo mass, and a Gaussian prior in a range of $1\leq c_{200}\leq20$ for the concentration with the median and scatter at $z=2$ from~\cite{2014MNRAS.441.3359D}. }

{Figure~\ref{fig:corner} shows the posterior distribution of $\sigma/m$, $M_{200}$, and $c_{200}$, where the contours denote $1\sigma$ and $2\sigma$ confidence regions inferred from the fit. Since there is only one datapoint from the observational constraints, i.e., the enclosed total mass within the Einstein radius $6.6~{\rm kpc}$, the halo mass is not well constrained for $M_{200}\gtrsim1.3\times10^{13}~{M_\odot}$. There is a trend that the cross section increases with the halo mass, but the correlation is not tightly constrained. Overall, the MCMC fit favors a small cross section of $\sigma/m\lesssim0.25~{\rm cm^2/g}$ within the $1\sigma$ confidence region ($\log_{10}\frac{\sigma/m}{\rm cm^2/g}\approx-0.6)$, consistent with the constraint $\lesssim0.3~{\rm cm^2/g}$ obtained in the main text.}

\bibliography{references}{}
\bibliographystyle{aasjournal}

\end{document}